\documentclass[3p,times,procedia]{elsarticle}
\usepackage{nupha_ecrc}

\volume{00}

\firstpage{1}

\journalname{Nuclear Physics A}

\runauth{}

\jid{nupha}

\jnltitlelogo{Nuclear Physics A}

\usepackage{amssymb}
\usepackage{amsmath}

\usepackage[figuresright]{rotating}

\begin{document}

\begin{frontmatter}



\dochead{}

\title{Longitudinal fluctuations and decorrelations of anisotropic flows in relativistic heavy-ion collisions}

\author[1]{Xiang-Yu Wu}
\author[2,3]{Long-Gang Pang}
\author[1]{Guang-You Qin}
\author[1,2,3]{Xin-Nian Wang}
\address[1]{Institute of Particle Physics and Key Laboratory of Quark and Lepton Physics (MOE), Central China Normal University, Wuhan, Hubei 430079, China}
\address[2]{Nuclear Science Division, Lawrence Berkeley National Laboratory, Berkeley, CA 94720, USA}
\address[3]{Physics Department, University of California, Berkeley, CA 94720, USA}

\begin{abstract}

We study the longitudinal decorrelations of elliptic, triangular and quadrangular flows in heavy-ion collisions at the LHC and RHIC energies.
The event-by-event CLVisc (3+1)-dimensional hydrodynamics model, combined with the fully fluctuating AMPT initial conditions, is utilized to simulate the space-time evolution of the strongly-coupled quark-gluon plasma. 
Detailed analysis is performed for the longitudinal decorrelations of flow vectors, flow magnitudes and flow orientations.
We find strong correlations between final-state longitudinal decorrelations of anisotropic flows and initial-state longitudinal structures and collision geometry:  
while the decorrelation of elliptic flow shows a non-monotonic centrality dependence due to initial elliptic geometry, 
typically the longitudinal flow decorrelations are larger in lower energy and less central collisions where the mean lengths of the string structure are shorter in the initial states.

\end{abstract}

\begin{keyword}
quark-gluon plasma \sep anisotropic flows \sep fluctuations \sep longitudinal decorrelations \sep relativistic hydrodynamics
\end{keyword}

\end{frontmatter}


\section{Introduction}
\label{}

Initial-state fluctuations and final-state anisotropic flows $\mathrm{V}_n$ provide important tools for studying strongly-coupled quark gluon plasma (QGP) in relativistic nuclear collisions at RHIC and the LHC.
By means of relativistic hydrodynamics simulation supplemented with fluctuating initial conditions, many theoretical studies have been focus on anisotropic flows and their fluctuations and correlations in the transverse directions. 
Recently, much interest has been paid on the fluctuations in the longitudinal (pseudorapidity) direction, which have the potential probe the longitudinal structure and evolution dynamics of the QGP fireball. 
Longitudinal fluctuations in the initial states imply that the final anisotropic flows may be different at two different rapidities: $\mathrm{V}_n(\eta_1) \neq \mathrm{V}_n(\eta_2)$ for $\eta_1 \neq \eta_2$.
In this work \cite{Wu:2018cpc}, we employ the CLVisc (ideal) (3+1)-dimensional hydrodynamics model \citep{Pang:2018zzo} combined with the AMPT \cite{Lin:2004en} initial conditions, and perform a systematic study on the longitudinal decorrelations of anisotropic flows, in terms of flow vectors, flow magnitudes and flow orientations, for different collision energies and centralities in heavy-ion collisions at RHIC and the LHC.

\begin{figure*}[thb]
\includegraphics[scale=0.32]{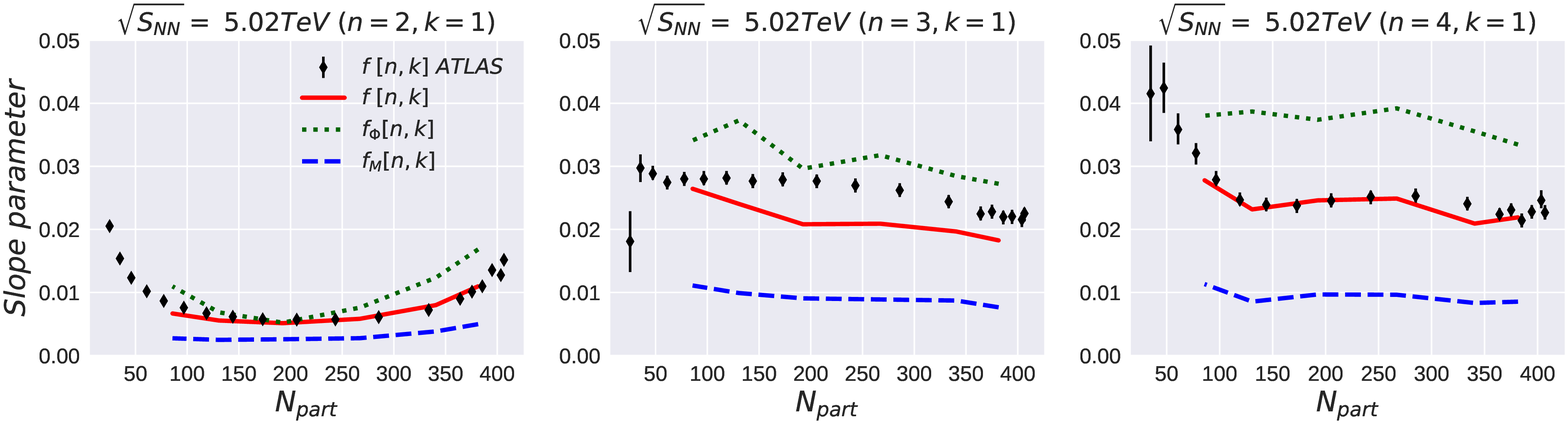}
\includegraphics[scale=0.32]{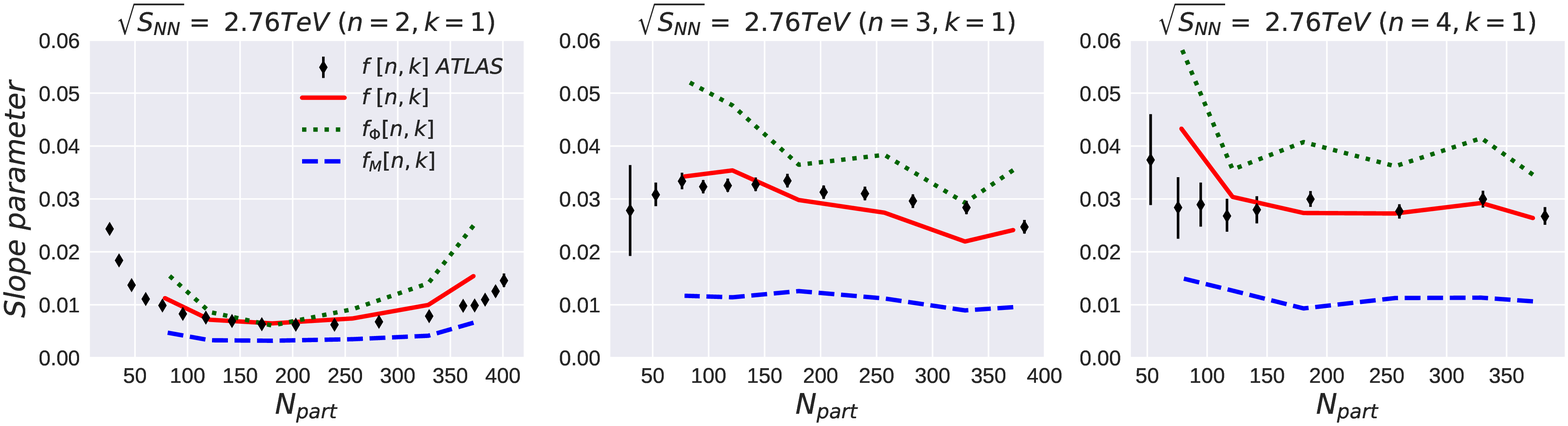}
\includegraphics[scale=0.32]{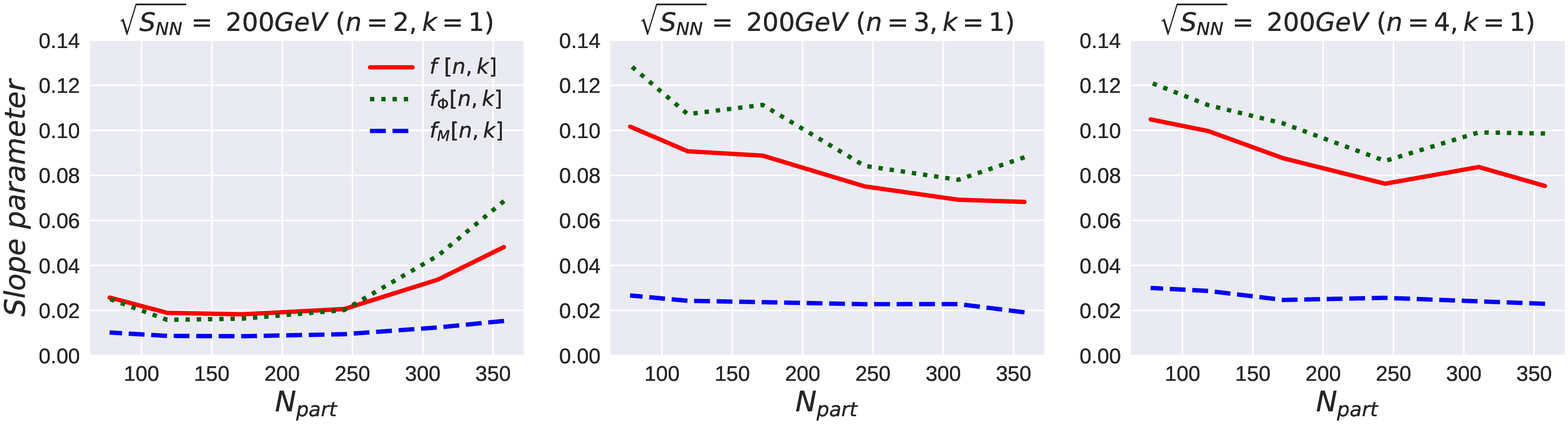}
	\caption{The slope parameters $f[n,k]$, $f_\Phi[n,k]$ and $f_M[n,k]$ for the longitudinal decorrelation functions $r[n,k](\eta)$, $r_\Phi[n,k](\eta)$ and $r_M[n,k](\eta)$ with $n$=2 (left) , $n$=3 (middle), $n$=4 (right) and $k$=1, as a function of collision centrality ($N_{\rm part}$) for Pb+Pb collisions at 5.02A TeV (upper), 2.76A TeV (middle) and for Au+Au collisions at 200A GeV (lower). 
	}
\label{coffr234}
\end{figure*}

\begin{figure*}[thb]
\includegraphics[scale=0.32]{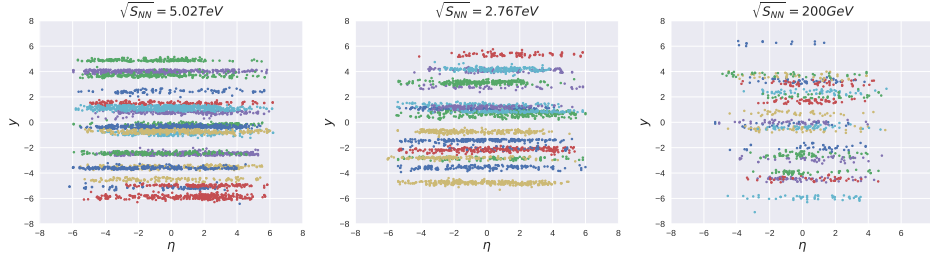}
	\caption{The string structures formed by the initial patrons in the AMPT model for typical events in Pb+Pb collisions at 5.02A TeV (left) and 2.76A TeV (middle) and for Au+Au collisions at 200A GeV (right).}
\label{length}
\end{figure*}

\begin{figure*}[thb]
\includegraphics[scale=0.32]{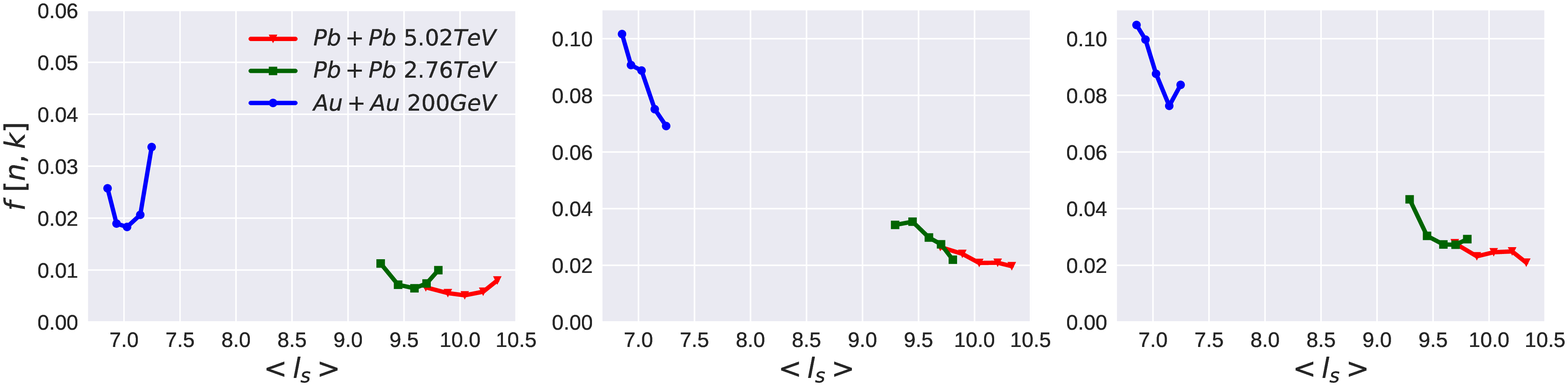}
\includegraphics[scale=0.32]{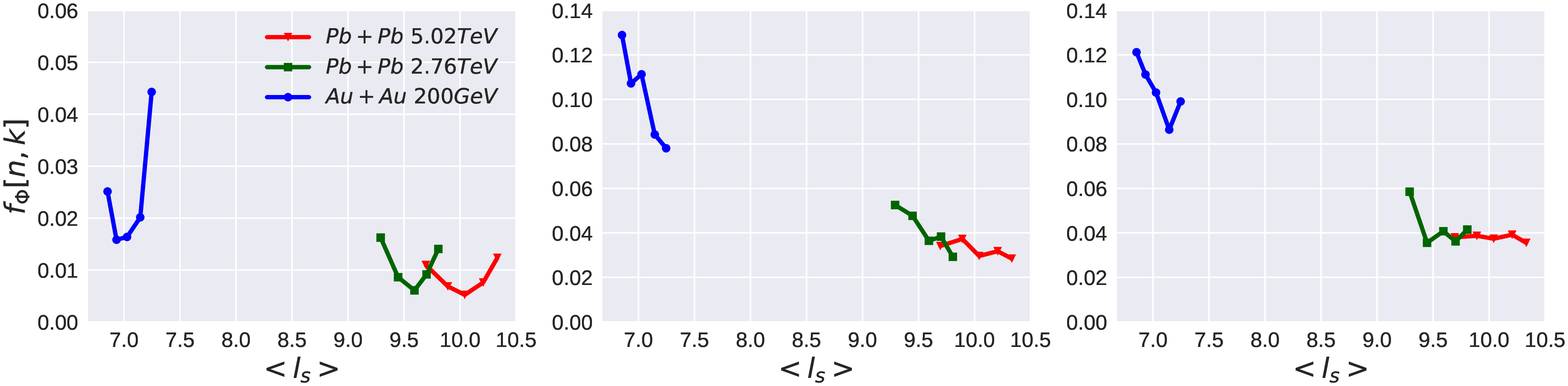}
\includegraphics[scale=0.32]{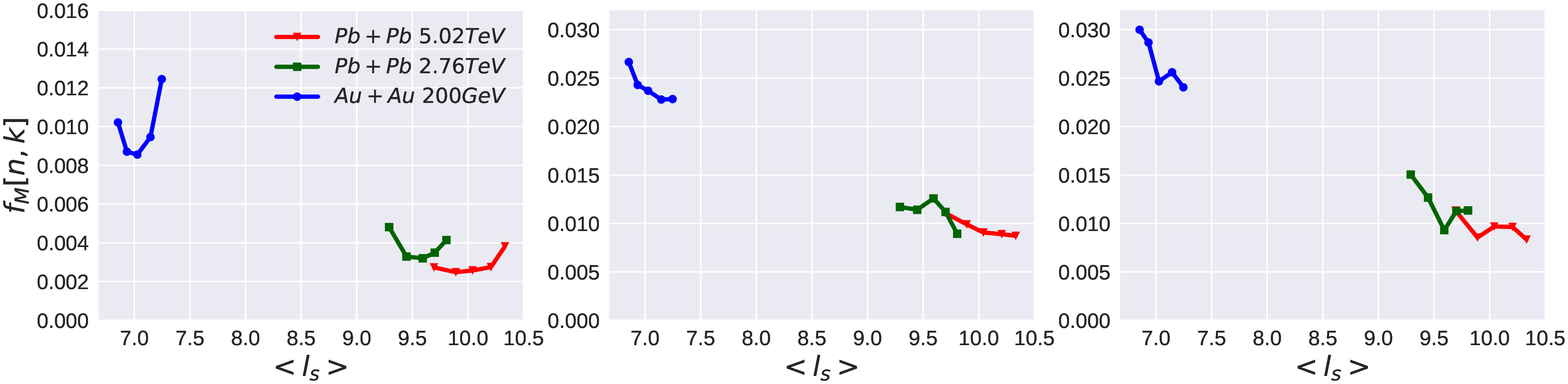}
	\caption{The slope parameters $f[n,k]$ (upper), $f_\Phi[n,k]$ (middle) and $f_M[n,k]$ (lower)  for the longitudinal decorrelation functions $r[n,k](\eta)$, $r_\Phi[n,k](\eta)$ and $r_M[n,k](\eta)$ 
	with $n$=2 (left), $n$=3 (middle), $n$=4 (right) and $k$=1 as function of collision energy and centrality (using the mean string length $\langle l_s \rangle$) for Pb+Pb collisions at 5.02A TeV and 2.76A TeV and for Au+Au collisions at 200A GeV.
In each panel, three curves denote three collision energies: 200A GeV, 2.76A TeV and 5.02A TeV, from left to right.
For each curve, five symbols represent five centrality bins: 40-50\%, 30-40\%, 20-30\%, 10-20\% and 5-10\%, from left to right.
	}
\label{mean}
\end{figure*}

\section{Event-by-event (3+1)-dimensional hydrodynamics simulation}

In this work, we use the ideal version of the CLVisc (3+1)-dimensional hydrodynamics model \citep{Pang:2018zzo} to simulate the dynamical evolution of the QGP fireball.
The initial conditions are obtained by the AMPT model \cite{Lin:2004en} which provides partons' positions and momenta for constructing the energy-momentum tensor:

\begin{eqnarray}
T^{\mu\nu} (\tau_0,x,y,\eta_s)= K \sum_i \frac{p^{\mu}_ip^{\nu}_i}{p^{\tau}_i}\frac{1}{\tau_0\sqrt{2\pi \sigma^2_{\eta_s}}}\frac{1}{2\pi \sigma^2_{r}}
 \exp\left[-\frac{(x-x_i)^2+(y-y_i)^2}{2\sigma_r^2}-\frac{(\eta_s-\eta_{si})^2}{2\sigma^2_{\eta_s}}\right] \,
\end{eqnarray}
The ideal hydrodynamic equation [$\partial_{\mu}T^{\mu\nu}=0, T^{\mu\nu}=U^{\mu}U^{\nu}(e+P)-Pg^{\mu\nu}$] is then numerically solved using KT algorithm.
The momentum distributions of final hadrons are computed using Copper-Frye formula, in which the freeze-out temperature is taken as $T_{fz}=137$MeV.

\section{Longitudinal decorrelations of anisotropic flows $\mathrm{V}_n$}
\label{res}

To study anisotropic flows, we use $\bf Q_n$-vector method: $\mathbf{Q}_n(\eta) = \frac{1}{N}\sum^N_{i=1} e^{in\phi_i} = q_n(\eta) \hat{Q}_n(\eta) = q_n(\eta) e^{in\Phi_n(\eta)}$,
where $q_n$ and $\Phi_n$ denote flow magnitude and flow orientation, respectively, in a given $\eta$ bin.
The longitudinal decorrelations of anisotropic flows may be studied using the following correlation functions \cite{Khachatryan:2015oea,Aaboud:2017tql,Jia:2014vja,Jia:2014ysa,Bozek:2017qir}: 
\begin{eqnarray}
r[n,k](\eta) &=& \frac{\langle \mathbf{Q}_n^k(-\eta) \mathbf{Q}_n^{*k}(\eta_{\rm r})\rangle}{\langle \mathbf{Q}_n^k(\eta)\mathbf{Q}_n^{*k}(\eta_{\rm r})\rangle} \,,
\quad
r_M[n,k](\eta) = \frac{\langle q_n^k(-\eta) q_n^{k}(\eta_{\rm r})\rangle}{\langle q_n^k(\eta) q_n^{k}(\eta_{\rm r})\rangle} \,,
\quad
r_{\Phi}[n,k](\eta) = \frac{\langle {\hat{Q}}_n^k(-\eta) {\hat{Q}}_n^{*k}(\eta_{\rm r})\rangle}{\langle {\hat{Q}}_n^k(\eta) {\hat{Q}}_n^{*k}(\eta_{\rm r})\rangle} ,
\label{origindef}
\end{eqnarray}
where $r[n,k](\eta)$, $r_M[n,k](\eta)$ and $r_{\Phi}[n,k](\eta)$ quantify the longitudinal decorrelations of flow vectors, flow magnitudes, and flow orientations, respectively.
Since the longitudinal decorrelation functions $r[n,k](\eta)$, $r_M[n,k](\eta)$ and $r_{\Phi}[n,k](\eta)$ decrease almost linearly as a function of pseudorapidity $\eta$, especially in small $\eta$ region \cite{Pang:2014pxa, Pang:2015zrq}, it is convenient to use the slopes of the decorrelation functions to quantify the longitudinal decorrelation effects. We follow ATLAS Collaboration \cite{Aaboud:2017tql} to define slope parameters as follows: 
\begin{eqnarray}
f[n,k] = \frac{\sum_i\left\{1 - r[n,k](\eta_i)\right\} \eta_i }{2\sum_i \eta_i^2} \,, 
f_M[n,k] = \frac{\sum_i\left\{1 - r_M[n,k](\eta_i)\right\} \eta_i }{2\sum_i \eta_i^2} \,,
f_\Phi[n,k] = \frac{\sum_i\left\{1 - r_\Phi[n,k](\eta_i)\right\} \eta_i }{2\sum_i \eta_i^2} \,.
\end{eqnarray}

Figure \ref{coffr234} shows the slope parameters $f[n,k]$, $f_\Phi[n,k]$ and $f_M[n,k]$ (with $k=1$), 
as function of collision centrality, for elliptic ($n$=2), triangular ($n$=3) and quadrangular ($n$=4) flows, for 2.76~TeV and 5.02~TeV Pb+Pb and 200~GeV Au+Au collisions. 
One can see that for all collision energies and centralities explored here, pure flow orientations have larger decorrelation effects than pure flow magnitudes, while the decorrelations of flow vectors sit in the middle. 
Also we observe a strong and non-monotonic centrality dependence for $\mathrm{V}_2$ decorrelation due to the initial-state elliptic collision geometry. 
Since $\mathrm{V}_3$ is dominated by fluctuations \cite{Qin:2010pf}, we observe weak centrality dependence for its longitudinal decorrelation, which tends to increase from central to less central collisions (similar trend for $\mathrm{V}_4$ decorrelation as well). 
Another important result is that $\mathrm{V}_n$ decorrelations are larger at RHIC (lower collision energy) than at the LHC (higher collision energies). 

The collision energy and centrality dependences of longitudinal decorrelations can be traced back to the longitudinal structures in the initial states. 
To illustrate this, we apply k-means algorithm to extract the lengths of the initial string structures in the AMPT model. 
The results are shown in Figure \ref{length}, for three typical central events in 200A GeV Au+Au collisions and 2.76A TeV, 5.02A TeV Pb+Pb collisions. 
We can see that the mean lengths of the initial string structures increase from lower to higher collision energies. 
Figure \ref{mean} shows the slope parameters $f[n,k]$, $f_\Phi[n,k]$ and $f_M[n,k]$ (with $k=1$) as function of collision energy and centrality (in terms of the mean string length) for $\mathrm{V}_2$, $\mathrm{V}_3$ and $\mathrm{V}_4$ at RHIC and the LHC energies. 
We find strong correlations between the longitudinal flow decorrelations in the final states and the longitudinal structures in the initial states: the longitudinal decorrelations of $\mathrm{V}_n$ tend to be  larger in lower energy and less central heavy-ion collisions where the lengths of the initial string structures are shorter.

\section{Summary}

In summary, we have performed a detailed analysis on the longitudinal decorrelations of elliptic, triangular and quadrangular flows in relativistic heavy-ion collisions at the LHC and RHIC energies. 
We find that the longitudinal decorrelation effects are much larger for pure flow orientations than for pure flow magnitudes. 
Also, a non-monotonic centrality dependence is found for the elliptic flow decorrelation.
Another important finding is that the longitudinal flow decorrelations are usually larger in lower energy and less central collisions due to the shorter lengths of the string structures in the initial states of heavy-ion collisions.

\section*{Acknowledgments}

This work is supported in part by NSFC (Nos. 11775095, 11375072 and 11221504), by the Major State Basic Research Development Program in China (No. 2014CB845404), by the Director, Office of Energy Research, Office of High Energy and Nuclear Physics, DNP, of the U.S. DOE (No. DE-AC02-05CH1123), and by the U.S. NSF within the framework of the JETSCAPE collaboration (No. ACI-1550228).




\bibliographystyle{elsarticle-num}
\bibliography{refs_GYQ}



\end{document}